\documentclass[structabstract]{aa}  
\usepackage{graphicx}
\usepackage{txfonts}
\usepackage[authoryear]{natbib}
\usepackage{url}
\usepackage{multirow}
\usepackage{array}

\begin{document}
   \title{Time evolution of high-energy emissions of low-mass stars:\thanks{
Based on observations collected at the Centro Astron\'omico Hispano Alem\'an
(CAHA) at Calar Alto, operated jointly by the Max-Planck Institut f\"ur
Astronomie and the Instituto de Astrof\'isica de Andaluc\'ia
(CSIC)},\thanks{Based on observations made with the WHT (William Herschel
Telescope) operated on the island of La Palma by the Isaac Newton Group in the
Spanish Observatorio del Roque de los Muchachos of the Instituto de
Astrof\'isica de Canarias.}}

   \subtitle{I. Age determination using stellar chronology with white dwarfs in
wide binaries}

\titlerunning{Age determination of low-mass stars with white dwarfs in
wide binaries}

   \author{A. Garc\'es,\inst{1}
           S. Catal\'an,\inst{2}
	   I. Ribas\inst{1}}
	%\fnmsep\thanks{Just to show the usage
        %  of the elements in the author field}
   \institute{Institut de Ci\`encies de l'Espai (IEEC-CSIC), Facultat de
Ci\`encies, Campus UAB, 08193 Bellaterra, Spain
%              \\
         \and
	    Centre for Astrophysics
Research, University of Hertfordshire, College Lane, Hatfield, AL10 9AB,
UK\\
	     \email{garces@ice.csic.es,
s.catalan@herts.ac.uk, iribas@ice.csic.es} }

   \date{Received 24 February 2011 / Accepted 30 March 2011}

% \abstract{}{}{}{}{} 
% 5 {} token are mandatory
 
  \abstract
   {Stellar ages are extremely difficult to determine and often subject to
large uncertainties, especially for field low-mass stars. We plan to carry out
a calibration of the decrease in high-energy emissions of low-mass GKM stars
with time, and therefore precise age determination is a key ingredient.  The
overall goal of our research is to study the time evolution of these high-energy
emissions as an essential input to studing exoplanetary
atmospheres.} 
   {We
propose to determine stellar ages with a methodology based on wide binaries. We are interested in
systems composed of a low-mass star and a white dwarf (WD), where the latter
serves as a stellar chronometer for the system. We aim at obtaining reliable
ages for a sample of late-type stars older than 1 Gyr. }
{We selected a sample of wide binaries composed by a DA type WD 
and a GKM companion. High signal-to-noise, low-resolution spectroscopic
observations were obtained for most of the WD members of the sample.
Atmospheric parameters were determined by fitting the spectroscopic data
to appropiate WD models. The total ages of the systems were derived by
using cooling sequences, an initial-final mass relationship and evolutionary
tracks, to account for the progenitor life.}
   {The spectroscopic observations have allowed us to determine ages for the
binary systems using WDs as cosmochronometers. We obtained reliable ages
for 27 stars  between 1 and 5 Gyr, which is a range where age determination
becomes difficult for field objects. Roughly half of these systems
have cooling ages that contribute at least 30\% the total age. We select those
for further study since their age estimate should be less prone to systematic
errors coming from the initial-final mass relationship.}
   {We have determined robust ages for a sizeable sample of GKM stars
that can be subsequently used to study the time evolution of their emissions
associated to stellar magnetic activity.}

   \keywords{stars:low-mass, stars: white dwarfs, stars: visual binaries,
stars: activity, stars: evolution}

   \maketitle
%
%________________________________________________________________

\section{Introduction}

The age of a star is one of the most difficult stellar parameters to determine.
Although several age-dating methods are becoming more precise, stellar
chronology still holds many uncertainties, especially when the objects are
in the field and moderately old. For ages above 0.7 Gyr, open clusters and
moving groups are, in general, not very useful as age calibrators, since they
are typically at large distances, making their individual component stars
difficult to study in detail, particularly their activity characteristics.
This is clearly illustrated in recently proposed calibrations \citep{Mamajek08}
that, while providing good performance for young ages (15--20\% uncertainty),
do not yield accurate results beyond 0.5 Gyr because of the lack of calibrators
and the increasing unreliability of rotation period estimates (differential
rotation effects or a weak signal from low-amplitude modulations). Furthermore, the
calibrations are not applicable to stars later than K2.

The determination of the ages of low-mass stars has many applications in
astrophysics \citep{Mamajek08,soderblom10}, including the study of galactic
populations or any use of intermediate-age G-, K-, and M- type stars as tracers
of evolution, such as calibrating of the decrease in high-energy emissions. Our interest 
lies in this use of age calibrations for
low-mass stars, because of the importance of stellar emissions, and their
time variability, to the understanding of exoplanet atmospheres. The host star
to a planet is, by far, its main source of energy, and many studies
\citep[e.g.][]{lamm03,Penz08} have demonstrated that stellar high-energy
emissions have a strong impact on planetary atmospheres. Thus, characterizing
such emissions is central to proper modelling of exoplanet
properties. High-energy emissions are related to the magnetic activity of the star,
and this activity decays throughout the stellar lifetime in an as yet poorly
understood manner, making it difficult to calculate the accumulated effects of
the UV and X-ray emission on the planetary environments. The decay in activity
with stellar age is intimately linked to the rotational evolution of the stars
\citep[e.g.][]{sku72,ayres97}.
Although the time evolution of high-energy emissions of solar-type stars is
already well constrained from \textit{The Sun in Time} project \citep{Ribas05},
that of late GKM stars still needs much improvement. Our objective is to extend \textit{The Sun in Time} 
project to cooler stars. This is
justified by the many differences in the high-energy emissions between, e.g.,
G-type and M-type stars and by the interest of the latter as hosts to
exoplanets. 

In this work we discuss the first part of our larger plan to determine
age-activity relationships for low-mass stars in general.  The paper is
organised as follows. In $\S$2 we present the results for stars with
stellar ages below 0.7 Gyr. The methodology of the age determination for WDs is
explained in $\S$3. In $\S$4 we collect our sample and describe its selection
criteria. In $\S$5 we present the observations and the data reduction.  Section
6 is devoted to the analysis, atmospheric parameter determination, and the
subsequent age determination method. This is followed by $\S$7 where we discuss
the low-mass companions in the context of defining age sequences, and finally
in $\S$8 we elaborate on our conclusions.
 
\section{Preliminary work}

We started by collecting data on the evolution of X-ray emissions for the
younger stars ($<$1 Gyr). Their ages were determined from cluster and moving
group membership \citep{mon01}, and the $\log L\rm_X$ values were obtained from a thorough
list provided by \cite{Pizzo03}. We complemented the measurements with X-ray
data estimated directly from ROSAT measurements for some field stars with age
estimates from different methods. 
 
The young main sequence phase was studied by means of stars belonging to the
IC2602, IC2391, Pleiades, $\alpha$\ Persei, Hyades clusters, and the Ursa
Majoris moving group. The early G-type star sample, as we explained in section 1, was
already available from \textit{The Sun in Time} project. These stars have
well-known rotation periods, temperatures, and metallicities. In general, we 
only consider stars in narrow spectral type bins (or effective temperature) to
avoid contamination from the intrinsic variation in magnetic activity with
stellar mass. The age range $<$0.07 Gyr is covered by stars in the IC2602, IC2391, 
and $\alpha$\ Persei clusters. We should point out that these clusters are
younger than 0.07 Gyr, (0.03, 0.03, and 0.05 Gyr, respectively), but since
saturation is present until 0.07 Gyr or longer in the case of K and M stars
\citep{ste01,jef10}, considering data with different ages
does not affect the value of $\log L\rm_X$.

For ages above 1 Gyr cluster or moving group membership
is not a useful age determination method. Other age indicators, such as the use
of rotation period, age-activity relations, asteroseismology, or theoretical
isochrones, are more useful in this age domain. Some of them have been used to
obtain the ages of a few GKM stars older than $\sim$1 Gyr. Only a handful of
field stars have reliable ages in this domain. These stars are 
$\alpha$~Cen~B and Proxima Cen, with ages
determined from the isochrone and asteroseismologic age of their close companion
$\alpha$~Cen~A \citep{Porto_de_mello08}. HR7703 shows space motions that
are typical of a thick disk star, so we can very roughly assume an age of
10 Gyr. The X-ray luminosities of these stars were determined from the ROSAT
database following the calibration in \cite{sch95}.For stars older
than $\sim$6 Gyr long-term changes in high-energy emissions are, in general,
difficult to distinguish from short-term stellar activity variations.

\begin{figure}[!t]
\centering
\includegraphics[angle=0,width=0.4\textwidth,clip]{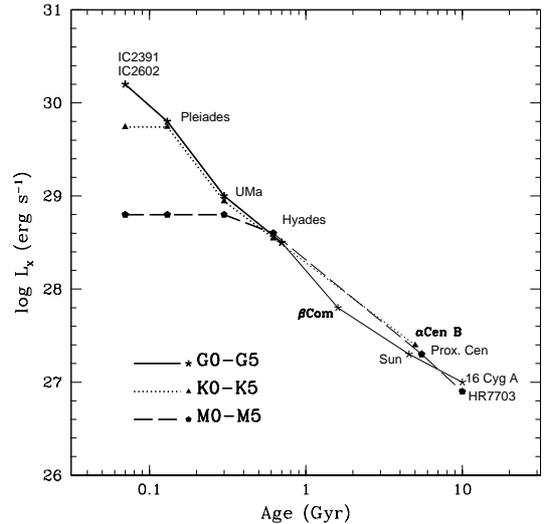}
\caption{X-ray luminosity versus age for solar-like stars and for field and
cluster K- and M-type stars. Only a few K and M stars are available beyond 0.7
Gyr.} 
\label{lx_age}
\end{figure}

We have put together all the compilled data, and the evolution of $\log L\rm_X$ with
age for three spectral type intervals (G0-5, K0-5, and M0-5) is illustrated in
Fig.~\ref{lx_age}.  In the case of G type stars, the plotted values are quite
reliable as they come from a thorough analysis of the \textit{The Sun in
Time} sample. For K- and M-type stars, however, the plotted values are
very crude results corresponding to the few stars described above that just
have rough age estimates. It is likely that the uncertainty of each point is at
least of a few tenths of a dex. The figure shows, as expected, that M-type
stars stay at saturated activity levels for a longer period of time than G-type
stars. According to these results, solar-like G0-5 stars are at saturated
emission levels until ages of $\sim$100 Myr, and their X-ray luminosity
decreases rapidly. K-type stars have saturated emission levels for a little
longer ($\sim$200 Myr) and then also decrease rapidly. Finally, M0-M5 stars
seem to have saturated emission levels up to 0.5 Gyr or more and then decrease
in an analogous way to G- and K-type stars.

The interval corresponding to ages youger than $\sim$0.7 Gyr is covered well by
the used cluster and moving group stars (IC2602, IC2391, Pleiades, $\alpha$\ Persei, Hyades clusters, and the Ursa
Majoris moving group), but it is obvious from our
preliminary analysis that a more complete sample of older stars is
needed to define reliable age-activity relationships. Furthermore, it is
interesting to note that an age of 0.7 Gyr is a key point for our Sun, since it
represents the time at which life is supposed to have appeared on the Earth's
surface. Thus, data for older stars of different types will be very important
for modelling planetary atmospheres in a regime that can be relevant to potential
life on their surface, as happened to our Earth. 

\section{Methodology}

We have developed an age determination method based on the use of wide binaries
where one of the components, a WD, is used as a chronometer. The members of a
wide binary are assumed to have been born simultaneously and with the same
chemical composition. Since they are well separated (100-1000 AU), we can
assume that no interaction has occurred between them in the past and they have
evolved as single stars \citep{Osw88}. We are interested in wide binaries
composed by a WD and a star with GKM spectral type. The evolution of a WD can
be described as a cooling process, which is relatively well understood at
present \citep{Sal00}. The total age of the WD can be expressed as the sum of
its cooling time plus the pre-WD lifetime of its progenitor. Thus, ages can be
obtained from an initial-final mass relationship and stellar tracks to account
for the pre-WD lifetime.  This procedure is analogous to that described by
\cite{cat08a}. It is sound to assume that the age of the WD is the same as the
that of the low-mass companion, since both members of a wide binary were born
simultaneously.

We selected a sample of 30 wide binaries containing a WD and a GKM star
(see Table~\ref{sample}). In the sample we favour WD components classified as a
DA, i.e., with the unique presence of Balmer lines in their spectra. As we demonstrate
in section 6, the fits to the spectral features of these WDs yield realistic values
for the atmospheric parameters (effective temperature, T$_{\rm eff}$ and
surface gravity, $\log g$). We have collected a large amount of observational
data (photometry and spectroscopy) on the WD components to implement the
proposed approach. We aim at determining total system ages (i.e., cooling ages
plus progenitor lifetime) with precisions of 10--20\%. This is sufficient in
our context since magnetic activity is an intrinsically variable phenomenon and
any relationship will have some inherent dispersion no matter how accurate the
ages and coronal/chromospheric fluxes.

\section{Sample selection}

The sample of wide binaries was compilled from the available literature. For
this we considered the recent revision of the New Luyten Two Tenths (NLTT) Catalogue 
\defcitealias{luy79}{Luyten 1979-1980}\citepalias{luy79} by
\cite{Chaname04}, an NLTT catalogue with Hipparcos stars from \cite{Gould04} and
later completed with some pairs from a selection of WD+M systems made by
\cite{Silvestri05}. 

Our first selection was done only considering the catalogues of
\cite{Chaname04} and \cite{Gould04} and following a careful evaluation
procedure.  First, the WD component had to be classified as a DA (i.e., with
the sole presence of Balmer lines), so that the fitting procedure can be
sufficiently accurate to derive atmospheric parameters. Secondly, the other
component of the pair should be a star of spectral type G, K, or M. For the WD
type classification we have used the \cite{McCook99,mcc06} catalogues, and the
spectral classification of the low mass stars was made by calculating their
\textit{V-J} index and taking the spectral classification into account for main
sequence stars in \cite{Johnson66}.  Information about the two proper motion
components of both members of the wide binary is available in \cite{Chaname04}
and \cite{Gould04} catalogues for most of the members of the sample. We have
completed and checked this information with data from USNO catalogue and
calculated the total proper motion of each star. We only consider in our sample
the wide binaries that satisfy the NLTT proper motion threshold, $\mu\geq$180
mas yr$^{-1}$ and that they show the same direction in their proper motions. 

The resulting sample was completed considering some wide binaries from the
catalogue of \cite{Silvestri05}. They studied the relationship between age and
chromospheric activity for 189 binary systems composed of a WD and an M-type
star. For this purpose, the authors determined the system's age using WDs as
stellar clocks but assuming a rough value for their mass, the typical 
0.6M$_{\odot}$ \citep{sil01}. Conversely, the ages estimated are potentially 
unreliable since the mass of each WD is an important parameter
for estimating their progenitor lifetime. With the aim of improving age estimates,
we performed a revision of the sample of \cite{Silvestri05} and completed our
sample by considering the pairs containing a DA WD and a GKM star in their
catalogue. The targets from \cite{Silvestri05} were also screened following the
proper motion criteria explained in the previous paragraph.

\begin{figure}[!t]
\centering
\includegraphics[angle=270,width=0.4\textwidth,clip]{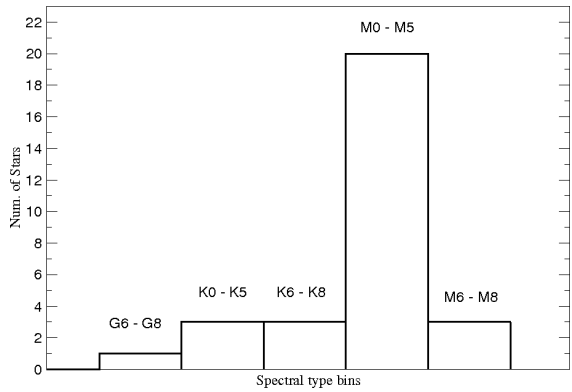}
\caption{Histogram of the spectral type distribution of the low-mass stars in
our sample, which are mostly of spectral type M.}
\label{histo}
\end{figure}

Our final sample is composed of 30 wide binaries. The sample with all relevant
information, including the \textit{V-J} indices of the companions and their
inferred spectral types, is provided in Table~\ref{sample}. For NLTT55287
\textit{J} photometry was not available and the spectral type was determined
from the \textit{B-V} index. In Table~\ref{sample} we provide an approximate
value for \textit{V-J} corresponding to an M8 star. Although photometric
information about the two members of the binary was available in the
catalogues, we checked all $V$ magnitudes in the SIMBAD database. We found some
differences or confusions between the magnitudes of the two members in some
cases. For these targets we made a thorough analysis to make sure that each
member was correctly identified. Figure~\ref{histo} shows a histogram with the
spectral type distribution of the low-mass stars. As can be seen, most of the
sample is composed by M-type stars plus a few K-type stars and one G-type
star. As we discuss in section 6, the WD sample mainly contains DA-type WDs, but also
a few DC-type WDs.

\begin{table}[!t]
\caption{The wide binary sample. Spectral types are estimated from the 
\textit{V-J} index, except NLTT55287, for which \textit{B-V} was used. 
}
\label{sample}
\centering
\small
\addtolength{\tabcolsep}{-1.5pt}
\begin{tabular}{lc|lcccc}
White dwarf & \textit{V} & Companion & \textit{V} & \textit{V-J} & SpT.  \\
\hline\hline
NLTT1762  &  16.59 &  NLTT1759  &  10.28 & 1.46 &  K1  \\
NLTT10976 &  17.20 &  NLTT10977 &  13.66 & 3.52 & M2 \\ 
NLTT13110 &  17.23 &  NLTT13109 &  12.63 & 3.21 &M2 \\ 
NLTT19311 &  16.58 &  NLTT19314 &  13.18 & 3.32 & M2 \\ 
NLTT21891 &  14.79 &  NLTT21892 &  15.41 & 4.22 & M5 \\ 
NLTT26379 &  12.92 &  NLTT26385 &  12.55 & 3.75 & M4 \\ 
NLTT28470 &  13.60 &  NLTT28469 &  14.09 & 4.13 & M5 \\
NLTT28712 &  15.55 &  NLTT28711 &  15.55 & 4.19 & M5 \\
NLTT28772 &  16.67 &  NLTT28771 &  16.72 & 4.18 & M5 \\
NLTT29967 &  17.26 &  NLTT29948 &  9.96  & 1.89 & K4 \\
NLTT31644 &  15.50  &  NLTT31647 &  16.89 & 5.06 & M6 \\
NLTT31890 &  15.86 &  NLTT31888 &  13.48 & 2.95 & M0 \\
NLTT39605 &  16.24 &  NLTT39608 &  15.74 & 3.85 & M4 \\
NLTT44348 &  17.50 &  NLTT44344 &  11.46 & 2.34 & K7 \\
NLTT56546 &  15.90  &  NLTT56548 &  16.60 & 3.99  & M4 \\
NLTT58107 &  16.13 &  NLTT58108 &  16.13 & 5.80: & M8 \\
G86-B1B   &  16.10 &  G86-B1A   &  14.26 & 3.67 & M4 \\
LP347-4   &  12.92 &  LP347-5   &  11.70 & 3.82 & M3 \\	
LP856-53  &  15.00 &  LP856-54  &  12.18 & 2.97 & M0 \\
LP888-64  &  13.56 &  LP888-63  &  13.80 & 4.23 & M5  \\
WOLF672A  &  14.34 &  WOLF672B  &  14.05 & 4.19 & M4  \\
%\hline
NLTT4615  &  17.48 &  NLTT4616  &  12.44 & 3.13 &M1 \\ 
NLTT7890  &  17.39 &  NLTT7887  &  9.84  & 1.90 &K3 \\ 
NLTT15796 &  17.21 &  NLTT15797 &  15.46 & 4.31 &M5 \\ 
G107-70   &  14.62 &  G107-69   &  13.52 & 4.38 &M6 \\
%\hline
NLTT1374 & 16.22   &   NLTT1370 &  12.90  &  2.19  &K6 \\
NLTT7051 & 16.18   &   NLTT7055 &  13.06 &  2.97 &M0 \\   
NLTT13599 &  15.94 &   NLTT13601 &  8.42  &  2.47    &K7 \\ 
NLT55288 & 16.50   &   NLTT55287&  8.03  &  1.21 &G7  \\
L577-71  & 12.80   &   L577-72  &  13.56 &  4.08 &M4  \\
\hline
\end{tabular}
\end{table}

\section{Observations and data reduction}

We performed optical long-slit low-resolution spectroscopic observations for
our analysis, with the goal of obtaining measurements with high signal-to-noise
ratio ($\sim$75--100). We were able to obtain spectroscopic data for 25
WDs in our sample during different observing campaigns between February of 2009
and February of 2010. The observing runs were carried out at Observatorio del
Roque de los Muchachos (ORM) and the German-Spanish Astronomical Center at
Calar Alto (CAHA). The 3.5m telescope was used during three nights in February of
2009 at CAHA, and the William Herschel Telescope (WHT) during 4.5 nights in
February and September of 2009 and February of 2010 at ORM. Table~\ref{instru}
provides more details about the observation setups used. We managed to observe
25 out of 30 initial targets, with the remaining five not visible at the time the
observing runs were allocated.

\begin{table}[!t]
 \caption{Information about the setup used for each observatory.
}
\addtolength{\tabcolsep}{-2pt}
\label{instru}
\centering
\small
\begin{tabular}{lcccccc}
Observ.& Telesc. &  Spectrogr. & R & Arm & Grism &Spectral      \\
       &         &             &   &     &       & coverage(\AA) \\
\hline\hline
\
CAHA   &  3.5m   & TWIN        & 1250 & Red &T04 &  5500$-$7500 \\
       &         &             &      &Blue &T08 &  3500$-$6500  \\
\\
ORM    &  WHT    & ISIS        & 2600 & Red & R316R & 4570$-$8430\\
       &         &             &      & Blue& R300B & 2730$-$6270\\
\hline
\end{tabular}
\end{table}

The spectroscopic observations covered the main H Balmer lines, from H$\alpha$
to H$\varepsilon$ or H9, whenever possible. The two spectrographs used included
two separate spectroscopic channels (blue and red arms) behind the common
entrance slit aperture. We chose a proper dichroic position in order to observe
the H$\alpha$ line in the red arm and all the other visible Balmer lines in the
blue arm, covering at least a spectral range of $\sim$3500--7500 \AA. A
suitable grism selection was done to place H$\alpha$ centred and unvignetted
for the red arm and a maximum number of unvignetted Balmer lines for the blue
arm. The slit width for each observation was chosen according to the seeing 
($\sim$1$^{\prime\prime}$--2$^{\prime\prime}$).
Spectra of high quality are essential to derive the atmospheric parameters with
accuracy. We performed as many exposures as necessary to guarantee a high
signal-to-noise ratio ($\sim$75--100) for the final spectrum of each object
(after the corresponding reduction). The exposure times and signal-to-noise
ratios for all the targets are shown in Table~\ref{exptime}. For targets with
individual observations with low signal-to-noise ratio, 
we co-added individual 1800 s exposures to minimise the effects of cosmic ray
hits on the CCD. 

\begin{table}[!t]
 \caption{Observation details for the targets in the sample.}
\centering
\begin{tabular}{lccc}
Observed    & Exp.       & S/N  & Num. of    \\
target	    & time (sec) &      & exposures  \\
\hline\hline
NLTT1762      & 9000 & 165 & 5 \\
NLTT4615      & 1800 & 20  & 1 \\
NLTT7890      & 3600 & 50  & 2 \\
NLTT10976     & 7200 & 75  & 4 \\ 
NLTT13110     & 3600 & 80  & 2 \\
NLTT15796     & 7200 & 105 & 4 \\
NLTT19311     & 12600& 160 & 7 \\
NLTT21891     & 5100   & 320 &5 \\
NLTT26379     & 1200   & 330 &5 \\
NLTT28470     & 1500   & 290 &5 \\ 
NLTT28712     & 6300   & 165 &5 \\
NLTT28772     & 13500  & 190 &9 \\
NLTT29967     & 12600  & 185 &7 \\
NLTT31644     & 7200   & 285 &6 \\
NLTT31890     & 7200   & 140 &4 \\
NLTT39605     & 7200   & 120 &4 \\
NLTT44348     & 7200   & 100 &4 \\
NLTT56546     & 3600   & 140 &2 \\ 
NLTT58107     & 3600   & 190 &2 \\ 
G86-B1B       & 11700  & 210 &8 \\
G107-70       & 3500   & 120 &3\\ 
LP347-4    & 500    & 130 &2 \\ 
LP856-53   & 4500   & 210 &5 \\
LP888-64   & 400    & 120 &2 \\
WOLF672A       & 4200   & 350 &6 \\
\hline
\end{tabular}
\label{exptime}
\end{table}

The WD spectra were reduced using the standard procedures within the
single-slit tasks of IRAF \footnote{IRAF is distributed by the National Optical
Astronomy observatory, which are operated by the Association of Universities
for Research in Astronomy, Inc., under cooperative agreement with the National
Science Foundation \url{http://iraf.noao.edu}}.
First, the images were bias-corrected and flatfield-normalised, and then, the
spectra were extracted and wavelength-calibrated using arc lamp observations.
Flux calibrations were not performed since they are not necessary for our
subsequent analysis, and most of the nights were not photometric. We applied
the heliocentric correction to each spectrum prior to coadding multiple
observations obtained in different nights.

\section{White dwarf analysis}

After the corresponding reduction, we carried out a first inspection of the
spectra. Although all the WDs we selected had previously been classified as DA,
we found that four of them do not show the bluer Balmer lines in their spectra, so
they are likely to be the DC type. For these four targets we checked the part of
the spectrum observed with the red arm. The H$\alpha$ line was not visible
in any of them, and is faint in DC WDs, the resolution of the 
instrument and the low signal-to-noise ratio
of the spectrum yields in the absence of such line for these four spectra. 
Table~\ref{sample} lists the observed 21 DA type WDs in the
first set of rows and remaining four observed DCs in the second set of rows. 

Once the spectra of the DA targets were correctly selected, we
proceeded to the continuum normalization procedure. The normalization 
has to be carried out carefully because the determination of 
atmospheric parameters is very sensitive to variations in the continuum. 
We carefully defined several clean continuum windows by considering the 
central part of each spectrum and avoiding any irregular feature or the 
wings of the spectral lines. We used Legendre polynomials of order 15 to 
20 for the normalization procedure.

\subsection{Atmospheric parameters}
 
Before calculating the atmospheric parameters of the WDs (T$_{\rm eff}$ and
$\log g$), we determined the radial velocities of each star using the IRAF task
{\sc fxcor}. Each spectrum was cross-correlated with a reference model from a
grid computed by D. Koester. The resulting radial velocity values can be
found in Table~\ref{atm_par}. The radial velocities determined were small in a
number of cases (ranging from 10 to 90~km~s$^{-1}$) when compared with the
resolution element of our observations. For some of the stars, however, the
measured radial velocities were higher (ranging from 100 to 200~km~s$^{-1}$)
thus becoming relevant to the subsequent analysis. In our procedure we took 
all the radial velocities for consistency into account.

We derived the atmospheric parameters of the WDs by performing a fit of the
observed Balmer lines using the models of D. Koester (private communication),
following a procedure based on $\chi^{2}$ minimization. The Balmer lines in
such WD models were calculated with the modified Stark broadening profiles of
\cite{Tremblay09}, kindly made available by the authors.  The model grid used
covers the range $T_{\rm eff}$=5000--100000 K, in steps of 1000 K at low
temperatures ($T_{\rm eff}$ $<$ 14000 K), steps of 2000 K (from $T_{\rm eff}$ =
16000 to 20000 K), steps of 5000 K (from $T_{\rm eff}$ = 20000 to 40000 K), and
steps of 10000 K for the higher temperatures ($T_{\rm eff}$ $>$ 40000 K).  The
range of $\log g$ is between 5.0 and 9.0 steps of 0.5 dex.  The WD models had
been previously normalised to the continuum and convolved with a Gaussian
instrumental profile with the proper FWHM in order to have the same resolution
as the observed spectra. The fit of the line profiles to synthetic models was
done using a subroutine based on the {\sc simplex} $\chi^{2}$ minimization
method of \cite{Press92}. We observed that the fitting procedure applied to all
the observed lines (in the red and blue arms) suffered from considerable
systematic effects. After running a number of tests we decided to consider only
the lines observed with the blue arm for the analysis to avoid systematic
errors because the H$\alpha$ line in the red arm is less adequate for
atmospheric parameter determination. 

Both $T_{\rm eff}$ and $\log g$ were obtained simultaneously. It is well known that
the changes in the profiles of the Balmer lines induce a certain degree of 
degeneracy to the procedure. There are often two numerically valid solutions
corresponding to minima of the $\chi^{2}$ function around the maximum
strength of the Balmer lines ($T_{\rm eff} \approx 12000$ K), i.e., a ``cool''
and a ``hot'' solution. In some cases these two minima are well separated
and their individual determination is straightforward. But for some other
objects, the parameters corresponding to the two scenarios are so close
as to become harder to distinguish. We tried to obtain both solutions for the
WDs by changing the initial $T_{\rm eff}$ for the analysis. We were able to
distinguish between the ``cool'' and ``hot'' solutions by finding the minimum
value of $\chi^{2}$ for both cases and comparing the results with the
photometric temperatures obtained as explained in section 6.2. 

\begin{figure}
\centering
\includegraphics[width=1\columnwidth,clip]{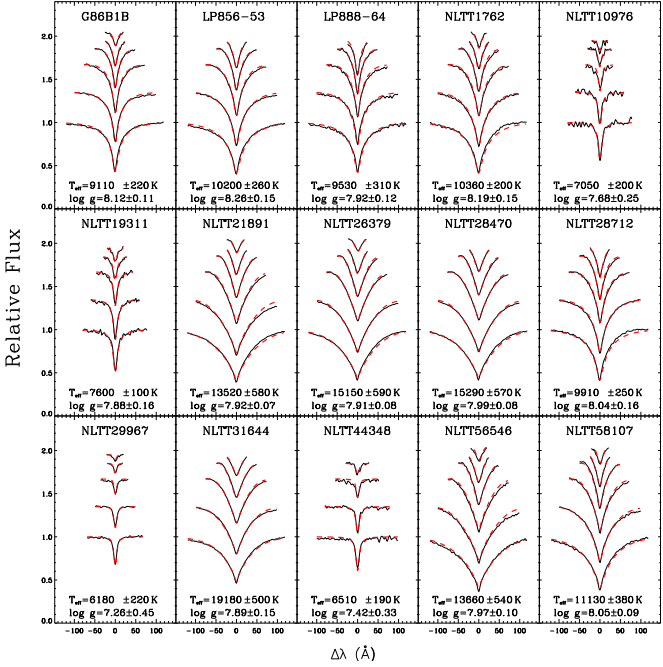}
\caption{Fits to the individual Balmer lines for the WDs observed at the WHT of
ORM. The line profiles correspond to H$\beta$ (\textit{bottom}) up to
H$\epsilon$, H8, or H9 (\textit{top}) depending on the object. We have applied
vertical shifts for clarity. The solid lines are the observed spectra and the
dashed lines the model that best fits the profile.}
\label{lp_fit_1}
\end{figure}

\begin{figure}
\centering
\includegraphics[width=1\columnwidth,clip]{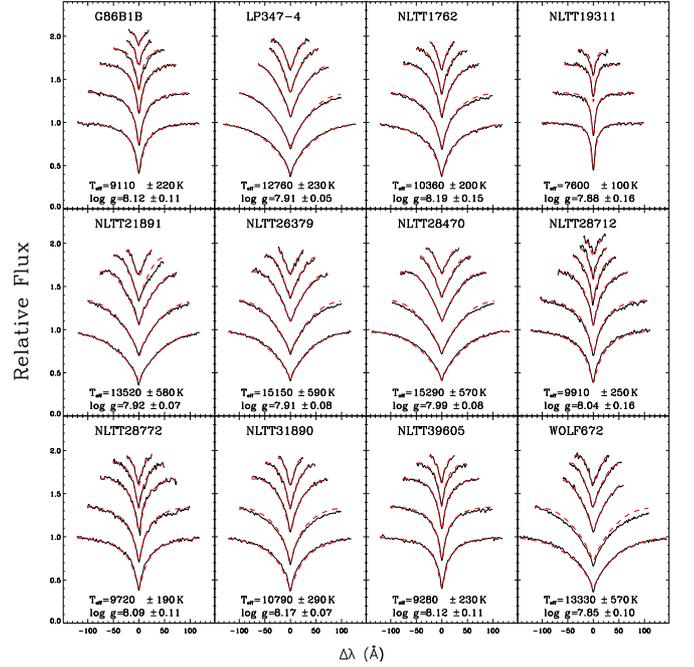}
\caption{Same as Fig. \ref{lp_fit_1} for the WDs observed at the 3.5-m
telescope at CAHA.}
\label{caha_fit}
\end{figure}

Given that our spectra have a very high signal-to-noise ratio, the main
contributor to the uncertainty in the fitting process (besides the adequacy of
the models themselves) is the normalization procedure. The normalization can
potentially modify the broad profiles of the spectral lines and thus introduce
systematic errors in the resulting parameters.  We carried out a careful
analysis to quantify such an error source. Using the IRAF {\sc continuum} task, we
obtained the blaze functions for each individual spectrum. To do so, 
we devised a procedure that is based on normalizing each individual spectrum
with the blaze functions corresponding to all the remaining spectra taken 
with the same instrument. Each of the resulting spectra was then re-normalised
(to correct for obvious trends) employing a 5-order Legendre function for the
observations with higher signal-to-noise and order 3 for the ones with low
signal-to-noise ratio. Finally, each of the resulting spectra for every target
were fitted to models as explained above to determine the atmospheric
parameters. The final uncertainties were estimated following the 
prescription of \cite{Ber92}, i.e., by deriving them from the independent fits
of the individual exposures for any given star before combination, but in our
case by considering all the normalised individual spectra. We are aware that
this procedure may overestimate the error since some of the renormalised
spectra were clearly not optimal. However, we prefer to make certain that all
possible systematic errors in the fitting procedure are accounted for by taking
a conservative approach and considering the uncertainties derived in this way.
The results of the fits and the estimated uncertainties are shown in
Table~\ref{atm_par}. Figures~\ref{lp_fit_1} and \ref{caha_fit} show the
fits to the Balmer line profiles for the DA WDs in our sample. 

\begin{table}[!t]
%\addtolength{\tabcolsep}{-6pt}
 \caption{Atmospheric parameters determined from fits to the Balmer lines to the observed WDs of our sample
and radial velocity values obtained for then.
 }
\label{atm_par}
\centering 
\small 
\begin{tabular}{lccc} 
WD Name & T$_{\rm eff}$ (K)  & log \textit{g} (dex) & Radial  \\ 
        &                    &                      &velocity (km s$^{-1}$) \\ 
\hline\hline
NLTT1762   &  10360$\pm$200 &  8.19$\pm$0.15 & 185$\pm$30\\  
NLTT13110  &   6630$\pm$220 &  7.99$\pm$0.19 & 210$\pm$15 \\ 
NLTT19311  &   7600$\pm$90  &  7.88$\pm$0.16 & 30$\pm$15\\
NLTT21891  &  13520$\pm$580 &  7.92$\pm$0.07 & 110$\pm$40 \\
NLTT26379  &  15150$\pm$590 &  7.91$\pm$0.08 & 110$\pm$40 \\
NLTT28470  &  15290$\pm$570 &  7.99$\pm$0.08 & 20$\pm$5\\ 
NLTT28712  &   9910$\pm$250 &  8.04$\pm$0.16 & 125$\pm$30\\
NLTT28772  &   9720$\pm$190 &  8.09$\pm$0.11 & 85$\pm$20\\
NLTT31644  &  19180$\pm$500 &  7.89$\pm$0.15 & 160$\pm$40\\
NLTT31890  &  10790$\pm$290 &  8.17$\pm$0.07 & 5$\pm$5\\
NLTT39605  &   9280$\pm$230 &  8.12$\pm$0.11 & 5$\pm$5\\
NLTT56546  &  13660$\pm$540 &  7.97$\pm$0.10 & 50$\pm$10\\ 
NLTT58107  &  11130$\pm$380 &  8.05$\pm$0.09 & 50$\pm$10\\ 
G86-B1B    &   9100$\pm$220 &  8.12$\pm$0.11 & 160$\pm$20\\ 
LP347-4    &  12760$\pm$230 &  7.91$\pm$0.05 & 90$\pm$30\\ 
LP856-53   &  10200$\pm$260 &  8.26$\pm$0.15 & 165$\pm$30 \\
LP888-64   &   9530$\pm$310 &  7.92$\pm$0.12 & 50$\pm$25 \\
WOLF672A   &  13330$\pm$570 &  7.85$\pm$0.10 & 15$\pm$5 \\
\hline
NLTT10976  &   7050$\pm$190 &  7.68$\pm$0.25 & 115$\pm$20\\ 
NLTT29967  &   6180$\pm$220 &  7.26$\pm$0.45 & 170$\pm$40\\
NLTT44348  &   6510$\pm$190 &  7.42$\pm$0.33 & $-$40$\pm$10\\
\hline
\end{tabular}
\tablefoot{The last three rows correspond to objects
for which the fit is poor.}
\end{table}

\subsection{Photometric effective temperatures}

Some stars in our sample were not visible in the epochs
of our observing runs. For these stars, and for the ones with no Balmer lines
in their spectrum, we used the available IR photometry to obtain their
photometric $T_{\rm eff}$. The $V$ magnitudes were obtained from the
catalogues used to build the sample and from the SIMBAD database, \textit{JHK}
photometry was collected from the 2MASS All Sky Catalog when available. For
$V$ magnitude we assumed an uncertainty of 0.02 mag as errors are seldom
available.

The $T_{\rm eff}$ for each object was obtained by adapting the method
of \cite{mas06} to WDs. This procedure consists in calculating synthetic
photometry using the WD atmospheres of \cite{hol06}. Subsequently, we developed
a fitting algorithm to compare the observational and theorical values and
minimise the $\chi^{2}$ using the Levenberg-Marquardt method, and $\chi^{2}$ was
defined from the differences between the observed and synthetic $VJHK$
magnitudes. The uncertainties in the calculated photometric temperatures were
derived from the covariance matrix.  The $VJHK$ magnitudes and photometric
$T_{\rm eff}$ for the sample stars are provided in Table~\ref{VJHK_Teff}.

\begin{table*}[!t]
%\addtolength{\tabcolsep}{-6pt}
 \caption{$VJHK$ magnitudes and photometric $T_{\rm eff}$ derived for
the stars in our sample.}
\label{VJHK_Teff}
\centering 
\small 
\begin{tabular}{lccccc} 
WD        & $V$             & $J$                 & $H$               & $K$                 & $T_{\rm eff}$ (K)     \\ 
\hline\hline
NLTT10976 & 17.20$\pm$0.02&  16.801$\pm$0.150 & 15.205$\pm$0.202& 15.783$\pm$0.232  & 7220$\pm$340  \\
NLTT19311 & 16.58$\pm$0.02&  16.070$\pm$0.103 & 15.833$\pm$0.194& 15.583$\pm$0.220  & 7760$\pm$320  \\
NLTT21891 & 14.79$\pm$0.02&  14.962$\pm$0.041 & 15.127$\pm$0.055& 15.055$\pm$0.167  &12700$\pm$670  \\
NLTT26379 & 12.92$\pm$0.02&  13.405$\pm$0.026 & 13.445$\pm$0.030& 13.544$\pm$0.056  &18880$\pm$1210 \\
NLTT28470 & 13.60$\pm$0.02&  14.013$\pm$0.036 & 13.984$\pm$0.059& 14.026$\pm$0.082  &15640$\pm$1150 \\
NLTT28712 & 15.55$\pm$0.02&  15.591$\pm$0.057 & 15.483$\pm$0.100& 15.734$\pm$0.177  &10700$\pm$380  \\
NLTT29967 & 17.26$\pm$0.02&  16.018$\pm$0.087 & 15.684$\pm$0.114& 15.500$\pm$0.238  & 5620$\pm$160  \\
NLTT31890 & 15.86$\pm$0.02&  15.802$\pm$0.067 & 15.627$\pm$0.136& 15.721$\pm$0.217  &  9980$\pm$320 \\
NLTT39605 & 16.24$\pm$0.02&  16.052$\pm$0.075 & 15.722$\pm$0.122& 15.889$\pm$0.279  &  9100$\pm$290 \\
NLTT56546 & 15.90$\pm$0.02&  16.571$\pm$0.137 & 15.995$\pm$0.197& 15.522$\pm$0.264  &13640$\pm$1050 \\
G86-B1B   & 16.10$\pm$0.02&  16.038$\pm$0.078 & 15.565$\pm$0.114& 14.562$\pm$0.109  & 8598$\pm$240  \\
LP347-4   & 12.92$\pm$0.02&  13.171$\pm$0.029 & 13.195$\pm$0.037& 13.179$\pm$0.028  &12640$\pm$500  \\
LP856-53  & 15.00$\pm$0.02&  14.907$\pm$0.110 & 14.797$\pm$0.249& 14.754$\pm$0.227  & 9800$\pm$450  \\
LP888-64  & 13.56$\pm$0.02&  13.188$\pm$0.067 & 13.205$\pm$0.061& 13.174$\pm$0.098  & 8940$\pm$190  \\
WOLF672A  & 14.26$\pm$0.02&  14.603$\pm$0.056 & 14.534$\pm$0.070& 14.562$\pm$0.109  & 13600$\pm$960 \\
\hline
NLTT4615  & 17.48$\pm$0.02 & 16.466$\pm$0.098 & 16.245$\pm$0.193& 15.555$\pm$0.181  &5950$\pm$190 \\
NLTT7890  & 17.39$\pm$0.02 & 16.131$\pm$0.081 & 16.131$\pm$0.081& 16.048$\pm$0.165  &5600$\pm$160 \\
NLTT15796 & 17.23$\pm$0.02 & 15.712$\pm$0.057 & 15.461$\pm$0.078& 15.511$\pm$0.155  &5025$\pm$60  \\
G107-70   & 14.62$\pm$0.02 & 13.083$\pm$0.022 & 12.838$\pm$0.022& 12.756$\pm$0.025  &5200$\pm$60  \\
\hline 
NLTT1374  & 16.22$\pm$0.02 & 16.050$\pm$0.082 & 15.842$\pm$0.168& 15.676$\pm$0.248  &9290$\pm$320 \\
NLTT7051  & 16.18$\pm$0.02 & 15.568$\pm$0.059 & 15.392$\pm$0.124& 15.353$\pm$0.206  &7630$\pm$220 \\
NLTT13599 & 15.94$\pm$0.02 & 14.598$\pm$0.038 & 14.232$\pm$0.058& 14.136$\pm$0.069  &5400$\pm$80  \\
NLTT55288 & 16.50$\pm$0.02 & 15.629$\pm$0.067 & 15.279$\pm$0.086& 15.195$\pm$0.168  &6510$\pm$170 \\
L577-71   & 12.80$\pm$0.02 & 13.183$\pm$0.033 & 12.885$\pm$0.033& 12.794$\pm$0.034  &11310$\pm$300\\
\hline
\end{tabular} 
\tablefoot{The table is divided into three sections: the first
one contains the observed DA WDs, the second one lists the DC WDs, and the
third one includes the WDs that have no spectroscopic observations.}
\end{table*}

\begin{figure}[!t]
\centering
\includegraphics[width=0.45\textwidth,clip]{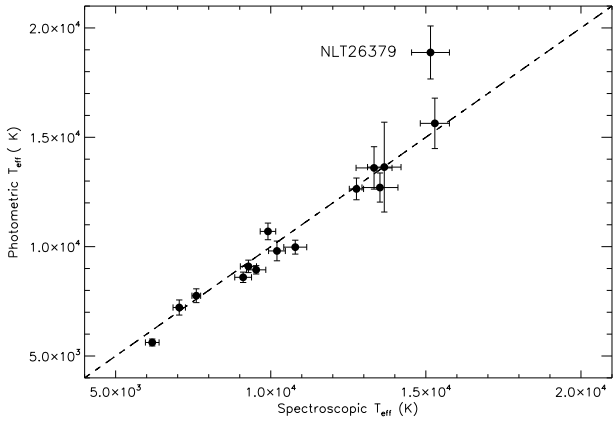}
\caption{Comparison of photometric and spectroscopic $T_{\rm eff}$ derived in
this work for 15 WDs in our sample.  The dashed line indicates 1:1
correspondence. The $\chi^2$ of the difference between the measurements is
0.57.}
\label{tsptph}
\end{figure}

In the cases where both spectroscopic data and $V$ and \textit{JHK} 2MASS
photometry were available, the photometric $T_{\rm eff}$ was used to evaluate
the reliability of the resulting $T_{\rm eff}$ obtained by the line profile
analysis. In most of the cases, both $T_{\rm eff}$ were in good agreement, as
shown in Fig.~\ref{tsptph}. NLTT26379 is the object showing the largest
deviation. For our subsequent analysis, we consider the $T_{\rm eff}$ value
obtained from spectroscopy, since it should be more reliable. It is worth
noting that the spectroscopic analysis of WDs with $T_{\rm eff}$ $<$ 12000K
should be done with special care, since their atmospheres could be enriched in
helium while retaining their DA spectral type \citep{Ber92}. In this case, photometric
$T_{\rm eff}$ provide the needed cross-check to ensure that the overall
atmospheric parameters are consistent and accurate.  It is reassuring to
confirm that our photometric and spectroscopic $T_{\rm eff}$ values are in good
concordance for the cooler targets of our sample.

Together with $T_{\rm eff}$, our procedure requires a value for $\log g$. 
Given the absence of any other possibility for the objects with photometric
$T_{\rm eff}$, we decided to adopt a ``mean'' $\log g$ estimated from the WDs
in the sample with spectroscopic parameters. We employed all WDs with reliable
$\log g$ values (i.e., the first part of Table \ref{atm_par}). We obtained
$\log g$ = 8.01$\pm$0.20, in which the error bar is not the standard deviation
but the overall range. 

\subsection{Masses, cooling times and total ages}

Once we derived the $T_{\rm eff}$ and $\log g$ of each WD, its mass ($M_{\rm
WD}$) and cooling time ($t_{\rm cool}$) can be obtained by using the
appropriate cooling sequences. We have adopted the cooling tracks of
\cite{Sal00}, which consider a carbon-oxygen (C/O) core WD (with a higher
abundance of O at the central core) with a thick hydrogen envelope ontop of a
helium buffer, $q({\rm H}) = M_{\rm H}/M = 10^{-4}$ and $q({\rm He}) = M_{He}/M
= 10^{-2}$. These improved cooling sequences include an accurate treatment of
the crystallization process of the C/O core, including phase separation upon
crystallization, together with up-to-date input physics suitable for computing
WD evolution. The resulting values for $M_{\rm WD}$ and $t_{\rm cool}$
are listed in Table \ref{results1}.

Systematic errors related to uncertainties on the CO stratification
and envelope composition may affect the cooling sequences \citep{sal09}. 
To check the sensitivity of our results to the adopted 
cooling tracks, we also used the sequences of \cite{fon01}, which consider different 
core and envelope compositions. The composition of the cooling sequences employed 
can be found in Table~\ref{cool_seq}.
 The derived masses do not change appreciably when adopting alternative cooling sequences,
with differences less than $\sim$0.02~M$_{\odot}$ over the entire mass range. 
The average of the differences in $t_{\rm cool}$ were found to be around 9$\%$. Thus, we
adopted a systematic error of 9$\%$ in $t_{\rm cool}$ in our subsequent analysis
of the cooling ages.

\begin{table}
 \caption{Cooling sequences, and layer composition.}
\begin{tabular}{l|c|c|c}
 Sequences    & Core & H envelope  & He buffer\\
	      &      & $q({\rm H})$&  $q({\rm He})$\\
\hline\hline
\cite{Sal00} & C/O       & 10$^{-4}$     & 10$^{-2}$\\
\hline
\cite{fon01}      & pure C    & 10$^{-4}$   & 10$^{-2}$\\
              & 50/50 C/O  & 10$^{-4}$   & 10$^{-2}$\\
	      & 50/50 C/O  & 10$^{-10}$  & 10$^{-2}$\\
\hline
\end{tabular}
\label{cool_seq}
\end{table}

 \begin{table*}
  \caption{Main results for the WDs in our sample.}
 \label{results1}
 \centering 
\addtolength{\tabcolsep}{2pt}
 \small
\setlength{\extrarowheight}{3pt}
 \begin{tabular}{lccccccccc}  
 WD      & $T_{\rm eff}$  & $\log g$ & $M_{\rm WD}$      & $t_{\rm cool}$ & $M_{\rm prog}$    & $t_{\rm prog}$ & Age &  $t_{\rm cool}$/Age & Comp \\                        
         &  (K)       &       & (M$_{\odot}$) & (Gyr)      & (M$_{\odot}$) & (Gyr)      & (Gyr)&  (\%) &\textit{V-J}\\
\hline\hline 
\textbf{NLTT1762}&\textbf{10360$\pm$200}&\textbf{8.19$\pm$0.15}&\textbf{0.72$\pm$0.07}&\textbf{0.68$\pm$0.10}&\textbf{2.96$\pm$0.52}&\textbf{0.49$_{-0.10}^{+0.11}$}&\textbf{1.2$_{-0.14}^{+0.15}$}&\textbf{58}&\textbf{1.46} \\
\textbf{NLTT13110}&\textbf{6630$\pm$220}&\textbf{7.99$\pm$0.19}&\textbf{0.59$\pm$0.08}&\textbf{1.61$\pm$0.35}&\textbf{1.64$\pm$0.87}&\textbf{2.3$_{-1.1}^{+2.6}$}&\textbf{3.9$_{-1.1 }^{+2.6}$}&\textbf{42}&\textbf{3.21}\\
NLTT19311&7600$\pm$100&7.88$\pm$0.16&0.52$\pm$0.07&0.94$\pm$0.13&0.96$\pm$0.73&12$_{-10}^{+8}$&13$_{-10}^{+8}$&7&3.32\\
NLTT21891&13520$\pm$580&7.92$\pm$0.07&0.57$\pm$0.03&0.23$\pm$0.03&1.45$\pm$0.32&3.4$_{-1.0}^{+1.1}$&3.6$_{-1.0}^{+1.1}$&6&4.22\\
NLTT26379&15150$\pm$590&7.91$\pm$0.08&0.56$\pm$0.03&0.16$\pm$0.02&1.41$\pm$0.36&3.8$_{-1.3}^{+1.4}$&3.9$_{-1.3}^{+1.4}$&4&3.75\\
NLTT28470&15290$\pm$570&7.99$\pm$0.08&0.61$\pm$0.03&0.18$\pm$0.02&1.86$\pm$0.38&1.49$_{-0.29}^{+0.47}$&1.67$_{-0.29}^{+0.47}$&11&4.13\\
\textbf{NLTT28712}&\textbf{9910$\pm$250}&\textbf{8.04$\pm$0.16}&\textbf{0.63$\pm$0.07}&\textbf{0.62$\pm$0.09}&\textbf{2.05$\pm$0.74}&\textbf{1.15$_{-0.34}^{+0.71}$}&\textbf{1.77$_{-0.35 }^{+0.71}$}&\textbf{35}&\textbf{4.19}\\
\textbf{NLTT28772}&\textbf{9720$\pm$190}&\textbf{8.09$\pm$0.11}&\textbf{0.65$\pm$0.05}&\textbf{0.69$\pm$0.08}&\textbf{2.32$\pm$0.55}&\textbf{0.89$_{-0.19}^{+0.26}$}&\textbf{1.58$_{-0.20}^{+0.27}$}&\textbf{44}&\textbf{4.18}\\
NLTT31644&19180$\pm$500&7.89$\pm$0.15&0.57$\pm$0.06&0.07$\pm$0.01&1.44$\pm$0.61&3.5$_{-1.7}^{+2.8}$&3.5$_{-1.7}^{+2.8}$&2&5.06\\
\textbf{NLTT31890}&\textbf{10790$\pm$290}&\textbf{8.17$\pm$0.07}&\textbf{0.71$\pm$0.03}&\textbf{0.60$\pm$0.06}&\textbf{2.88$\pm$0.29}&\textbf{0.53$_{-0.06}^{+0.07}$}&\textbf{1.13$_{-0.09}^{+0.09}$}&\textbf{53}&\textbf{2.95}\\
\textbf{NLTT39605}&\textbf{9280$\pm$230}&\textbf{8.12$\pm$0.11}&\textbf{0.67$\pm$0.05}&\textbf{0.82$\pm$0.10}&\textbf{2.52$\pm$0.55}&\textbf{0.76$_{-0.18 }^{+0.17}$}&\textbf{1.58$_{-0.20 }^{+0.19}$}&\textbf{52}&\textbf{3.85}\\
NLTT56546&13660$\pm$540&7.97$\pm$0.10&0.59$\pm$0.04&0.24$\pm$0.03&1.72$\pm$0.45&1.93$_{-0.54}^{+0.87}$&2.17$_{-0.54}^{+0.87}$&11&3.99\\
\textbf{NLTT58107}&\textbf{11130$\pm$380}&\textbf{8.05$\pm$0.09}&\textbf{0.64$\pm$0.04}&\textbf{0.47$\pm$0.05}&\textbf{2.15$\pm$0.45}&\textbf{1.04$_{-0.18 }^{+0.30}$}&\textbf{1.51$_{-0.19}^{+0.30}$}&\textbf{31}&\textbf{5.80:}\\
\textbf{G86-B1B}&\textbf{9110$\pm$220}&\textbf{8.12$\pm$0.11}&\textbf{0.67$\pm$0.05}&\textbf{0.86$\pm$0.10}&\textbf{2.51$\pm$0.56}&\textbf{0.77$_{-0.19}^{+0.18}$}&\textbf{1.64$_{-0.22}^{+0.21}$}&\textbf{53}&\textbf{3.67}\\
LP347-4&12760$\pm$230&7.91$\pm$0.05&0.56$\pm$0.02&0.27$\pm$0.03&1.36$\pm$0.26&4.3$_{-1.1}^{+1.0}$&4.5$_{-1.1}^{+1.0}$&6&3.82\\
\textbf{LP856-53}&\textbf{10200$\pm$260}&\textbf{8.26$\pm$0.15}&\textbf{0.76$\pm$0.07}&\textbf{0.80$\pm$0.14}&\textbf{3.26$\pm$0.53}&\textbf{0.37$_{-0.07}^{+0.08}$}&\textbf{1.17$_{-0.16}^{+0.16}$}&\textbf{68}&\textbf{2.97}\\
LP888-64&9530$\pm$310&7.92$\pm$0.12&0.56$\pm$0.05&0.58$\pm$0.07&1.34$\pm$0.52& 4.5$_{-2.1}^{+3.3}$&5.1$_{-2.1}^{+3.3}$&11&4.23\\ 
WOLF672A &13330$\pm$570&7.85$\pm$0.10&0.53$\pm$0.04&0.22$\pm$0.03&1.00$\pm$0.44&10$_{-5}^{+4}$&11$_{-5}^{+4}$&2&4.19\\ 
\hline
NLTT10976&  7050$\pm$200&7.68$\pm$0.25& 0.40$\pm$0.15&0.99$\pm$0.12 &-&-& -&-&3.52\\
NLTT29967&  6180$\pm$220&7.26$\pm$0.45& 0.14$\pm$0.19&0.55$\pm$0.23 &-&-& -&-&1.89\\
NLTT44348&  6510$\pm$190&7.42$\pm$0.33& 0.23$\pm$0.14&0.63$\pm$0.20 &-&-& -&-&2.34\\
\hline
\hline
\textbf{NLTT4615}&\textbf{5950$\pm$190}&\textbf{8.01$\pm$0.20}&\textbf{0.59$\pm$0.09}&\textbf{2.40$\pm$0.68}&\textbf{1.73$\pm$0.94}&\textbf{1.9$_{-0.9}^{+2.2}$}&\textbf{4.3$_{-1.1}^{+2.3}$}&\textbf{56}&\textbf{3.13}\\ 
\textbf{NLTT7890}&\textbf{5600$\pm$160}&\textbf{8.01$\pm$0.20}&\textbf{0.59$\pm$0.09}&\textbf{2.99$\pm$0.78}&\textbf{1.71$\pm$0.94}&\textbf{2.0$_{-0.9}^{+2.3}$}&\textbf{5.0$_{-1.2}^{+2.5}$}&\textbf{60}&\textbf{1.90}\\
\textbf{G107-70} &\textbf{5210$\pm$ 60}&\textbf{8.01$\pm$0.20}&\textbf{0.59$\pm$0.09}&\textbf{3.90$\pm$1.00}&\textbf{1.69$\pm$0.94}&\textbf{2.0$_{-1.0}^{+2.5}$}&\textbf{5.9$_{-1.4}^{+2.7}$}&\textbf{66}&\textbf{4.38}\\
\textbf{NLTT1374}&\textbf{9290$\pm$330}&\textbf{8.01$\pm$0.20}&\textbf{0.60$\pm$0.09}&\textbf{0.70$\pm$0.12}&\textbf{1.83$\pm$0.92}&\textbf{1.6$_{-0.6}^{+1.6}$}&\textbf{2.3$_{-0.7}^{+1.6}$}&\textbf{31}&\textbf{2.19}\\
\textbf{NLTT7051}&\textbf{7630$\pm$220}&\textbf{8.01$\pm$0.20}&\textbf{0.60$\pm$0.09}&\textbf{1.16$\pm$0.22}&\textbf{1.78$\pm$0.93}&\textbf{1.7$_{-0.7}^{+1.8}$}&\textbf{2.8$_{-0.8}^{+1.8}$}&\textbf{41}&\textbf{2.97}\\
\textbf{NLT13599}&\textbf{5400$\pm$ 80}&\textbf{8.01$\pm$0.20}&\textbf{0.59$\pm$0.09}&\textbf{3.41$\pm$0.87}&\textbf{1.70$\pm$0.95}&\textbf{2.0$_{-0.9}^{+2.4}$}&\textbf{5.4$_{-1.3}^{+2.6}$}&\textbf{63}&\textbf{2.47}\\
\textbf{NLTT15796}&\textbf{5025$\pm$60}&\textbf{8.01$\pm$0.20}&\textbf{0.59$\pm$0.09}&\textbf{4.60$\pm$0.88}&\textbf{1.68$\pm$0.95}&\textbf{2.1$_{-1.0}^{+2.6}$}&\textbf{6.7$_{-1.3}^{+2.8}$}&\textbf{69}&\textbf{4.31}\\
\textbf{NLT55288}&\textbf{6510$\pm$170}&\textbf{8.01$\pm$0.20}&\textbf{0.60$\pm$0.09}&\textbf{1.77$\pm$0.43}&\textbf{1.75$\pm$0.94}&\textbf{1.8$_{-0.8}^{+2.1}$}&\textbf{3.6$_{-0.9}^{+2.1}$}&\textbf{49}&\textbf{1.21}\\
\textbf{L577-71} &\textbf{8150$\pm$110}&\textbf{8.01$\pm$0.20}&\textbf{0.60$\pm$0.09}&\textbf{0.98$\pm$0.17}&\textbf{1.79$\pm$0.92}&\textbf{1.6$_{-0.7}^{+1.7}$}&\textbf{2.6$_{-0.7}^{+1.7}$}&\textbf{37}&\textbf{4.08}\\
\hline
\end{tabular}
\tablefoot{The WDs with $t_{\rm cool}$/Age ratio greater than 30\% are 
highlighted in boldface.}
\end{table*}

For three of our targets, included at the bottom of Table~\ref{atm_par}, our
spectroscopic fitting procedure yielded low values of $\log g$ ($<7.7$) with
relatively poor fits. The photometric temperatures available for two of these
targets (NLTT10976 and NLTT29967) show good agreement with the spectroscopic
value, thus indicating the consistency of the analysis. Thus, although the $\log g$
values that we obtained seem reliable, we could not infer reliable masses for
them using the cooling sequences of \cite{Sal00} since they are out of the
covered parameter range. The values for $M_{\rm WD}$ and $t_{\rm cool}$ obtained via extrapolation 
for these three targets can be found in Table~\ref{results1}. We obtained $M_{\rm WD}$ $<$ 0.5 M$_{\odot}$, 
which yields progenitor masses lower than 1 M$_{\odot}$.

From the derived $M_{\rm WD}$ and $t_{\rm cool}$ values, the mass of the
progenitor ($M_{\rm prog}$) can be determined using an initial-final mass relationship
\citep{cat08b}. From the calculated $M_{\rm prog}$, we used the stellar
tracks of \cite{dom99} to derive the progenitor's lifetime ($t_{\rm prog}$).
The total age of the WDs, and consequently that of the low-mass companion,
follows directly by adding $t_{\rm prog}$ and $t_{\rm cool}$. The final age
value is provided in Table~\ref{results1}. In this Table we compiled the final
atmospheric parameters, the mass of the WD, the mass of the progenitor star,
the cooling time of the WD, the main-sequence time of the progenitor 
star, the total age, the ratio between the cooling time and the total age, and
the \textit{V-J} colour of the low-mass companions. The first 18 WDs in the 
Table have
atmospheric parameters determined from the Balmer line profile fitting.  The three
objects in the middle were analysed spectroscopically, but the derived
$\log g$ are outside of the valid parameter range of our procedure. The nine WDs
at the end of the table have temperatures determined by $VJHK$ photometry and
the mean $\log g$ value for our sample estimated from the spectroscopic determinations.

\begin{figure}
\centering
\includegraphics[width=0.45\textwidth,clip]{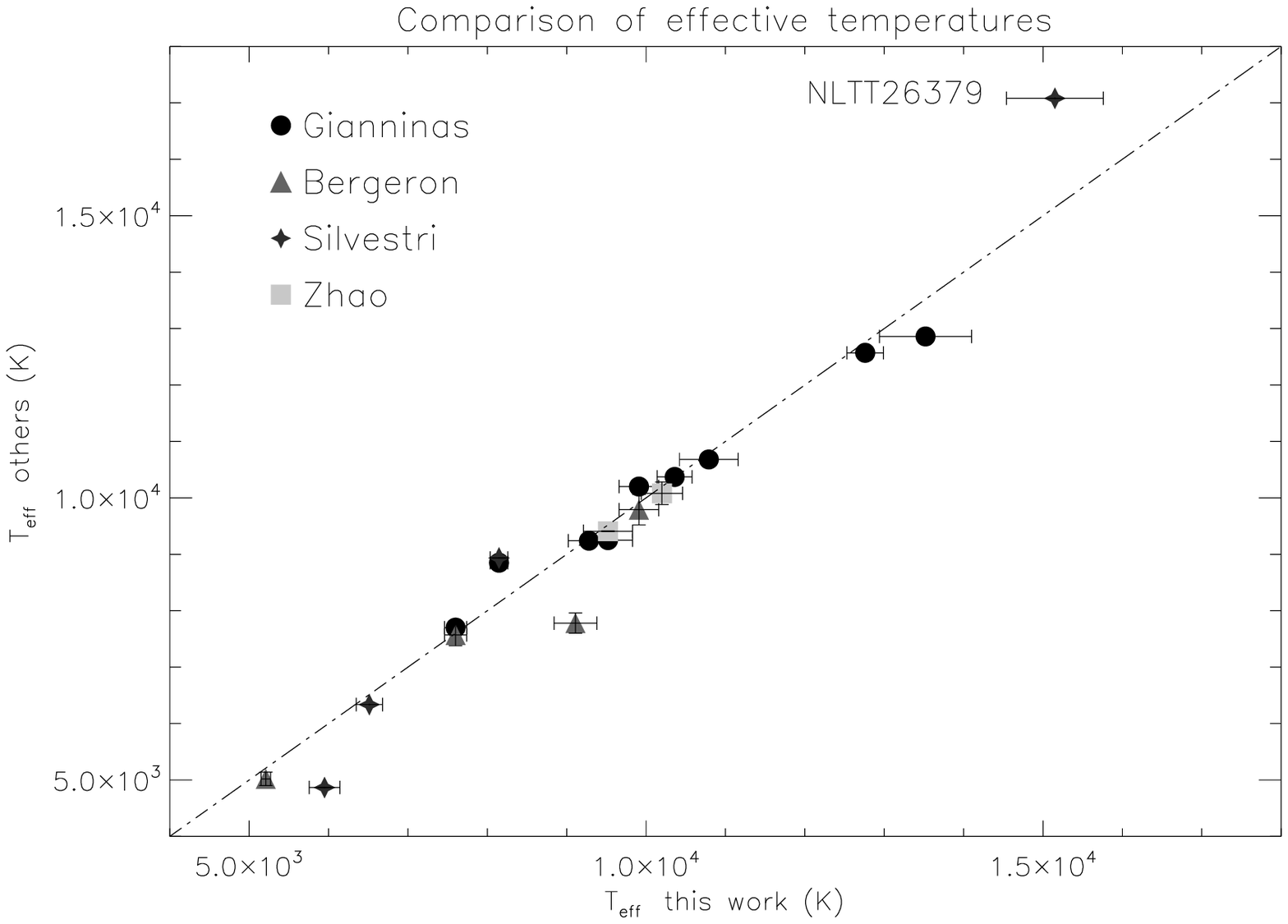}
\includegraphics[width=0.45\textwidth,clip]{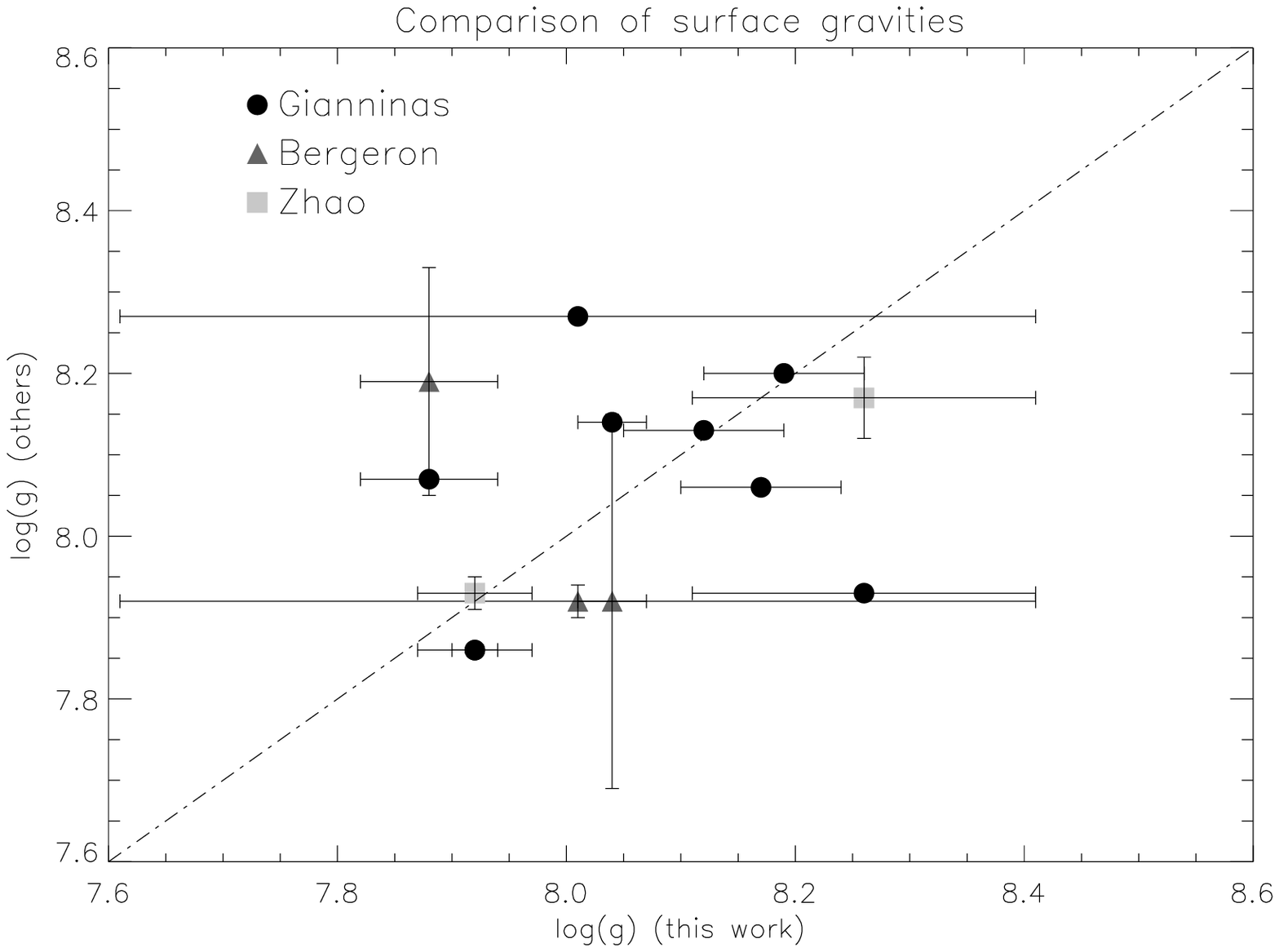}
\caption{Comparison of $T_{\rm eff}$ (top) and $\log g$ (bottom) determined by
us with values obtained by \cite{ber01}, \cite{sil01}, \cite{gianninas05}, and
\cite{Zhao11}. The dashed line is the 1:1 correspondence.}
\label{t_log_comp}
\end{figure}

Some of the WDs in our sample have atmospheric parameters and, in some cases,
masses, cooling times, and ages estimated from previous investigations, such as
\cite{ber01}, \cite{sil01}, \cite{gianninas05}, and \cite{Zhao11}. We can
compare our results to the previous determination. Our $T_{\rm eff}$ and $\log
g$ values are compared to those in the literature and are shown in
Fig.~\ref{t_log_comp}. In general the agreement between the $T_{\rm eff}$
determinations is good within the uncertainties, but there are some discrepant
results for some of the targets, such as NLTT26379 and G86-B1B. In the case of
G86-B1B, our analysis shows good accord between spectroscopic and photometric
determinations and thus, given the available cross-check, we prefer the value
we determined. In the case of NLTT26379, our photometric and spectroscopic
$T_{\rm eff}$s are discrepant at the 2.7-$\sigma$ level, and the result of
\cite{sil01} (using photometry) lies squarely in between. Given the better a
priori reliability of the line profile analysis for $T_{\rm eff}$ above 12000 K, we subsequently adopt the
spectroscopic $T_{\rm eff}$ value. The comparison of the $\log g$ values
between our values and the literature yields no very discrepant results. 

\subsection{Photometric distances}

We have collected available parallax measurements for the low-mass companions
and completed the information with photometric distances. The photometric
distances were determined using the evolutionary models for low-mass stars
described in \cite{baraffe97} and using the ages determined for each wide
binary system. The results for the sample are collected in Table~\ref{dist}. 
For NLTT19314 and WOLF672B, we could not directly perform
a photometric distance determination since their estimated ages are greater 
than 8 Gyr, which is the age limit of the
models we used. Thus, we just adopted a value of 8 Gyr, since the evolutionary
changes of these stars between 8 and 13 Gyr are negligible.

\begin{table}[!t]
\caption{Distances of the low-mass members of the wide binary sample.} 

\label{dist}
\centering
\small
\begin{tabular}{lc}
Star & $d$ (pc) \\
\hline\hline
NLTT1759  &105$\pm$15*    \\
NLTT10977 & 60$\pm$5      \\ 
NLTT13109 & 20$\pm$5*    \\ 
NLTT15797 & 40$\pm$5*     \\ 
NLTT19314 & 45$\pm$20     \\ 
NLTT21892 & 30$\pm$5*    \\ 
NLTT26385 & 15$\pm$5     \\ 
NLTT28469 & 15$\pm$5     \\
NLTT28711 & 55$\pm$15*    \\
NLTT28771 & 50$\pm$5     \\
NLTT29948 & 45$\pm$5*    \\
NLTT31647 & 25$\pm$5     \\
NLTT31888 & 80$\pm$5     \\
NLTT39608 & 60$\pm$5     \\
NLTT44344 & 55$\pm$5     \\
NLTT56548 & 60$\pm$5   \\
\hline
\end{tabular}
\hspace*{5mm}
\begin{tabular}{lc}
Star & $d$ (pc) \\
\hline\hline
NLTT58108 & 90$\pm$10       \\
G86-B1A   & 25$\pm$5  \\
LP347-5   &  8$\pm$5   \\	
LP856-54  & 45$\pm$5  \\
LP888-63  & 15$\pm$5  \\
WOLF672B  & 15$\pm$5   \\
NLTT4616  & 35$\pm$5  \\ 
NLTT7887  & 45$\pm$5  \\ 
G107-69   & 10$\pm$5  \\
NLTT1370 & 125$\pm$10  \\
NLTT7055 &  65$\pm$5   \\   
NLTT13601&  30$\pm$5*  \\ 
NLTT55287&  35$\pm$5* \\
L577-72  &  15$\pm$5  \\
\hline
&\\
\end{tabular}
\tablefoot{* nine
low-mass stars with trigonometric parallax information.}
\end{table}

\section{Discussion}

The main product of this part of the work is the determination of ages for
27 low-mass stars that are members of wide binary pairs. For eighteen of these age
determinations we employed WD atmospheric parameters coming from a
detailled spectroscopic fit, while for nine additional systems we used
photometric $T_{\rm eff}$ values and an estimated $\log g$ from the 
average of the rest. The final age is the sum of two contributions, namely
the cooling time of the WD and the lifetime of the progenitor, mostly during
the main sequence. As can be seen in Table~\ref{results1}, there is a clear
distinction between the subsamples with spectroscopic and photometric
temperatures in that the former shows relatively short cooling ages (usually
$<1$ Gyr), while these are significantly longer for the latter ($\sim$1--5 Gyr).
This is not a coincidence since older WDs are cooler and the Balmer lines
in their spectrum are weak or nonexistent. 

To further exploit our sample, it is worth evaluating the reliability of our
estimated ages. The uncertainties we provide in Table~\ref{results1} come 
from the random error estimates related to our fitting procedure, but they 
do not include systematic contributions potentially arising from the 
relationships and models used. We are not especially concerned with the 
contribution from models since the WD cooling process is relatively simple 
and main-sequence lifetime estimates should be quite reliable. However, 
there is some degree of concern as to the possible existence of systematic 
uncertainties
related to the initial-final mass relationship of WDs. Many
improvements have been achieved during these last years, for instance with the
coverage of the low-mass domain \citep{cat08b}. However, there is still a 
relatively large cosmic dispersion in the empirical initial-final mass
relationship that could be related, e.g., to metallicity effects or the 
poorly-known mass-loss processes during the AGB phase. For this reason, 
we believe that $t_{\rm cool}$ is more reliable than $t_{\rm prog}$. 

\begin{figure}[!t]
\centering
\includegraphics[width=0.45\textwidth,clip]{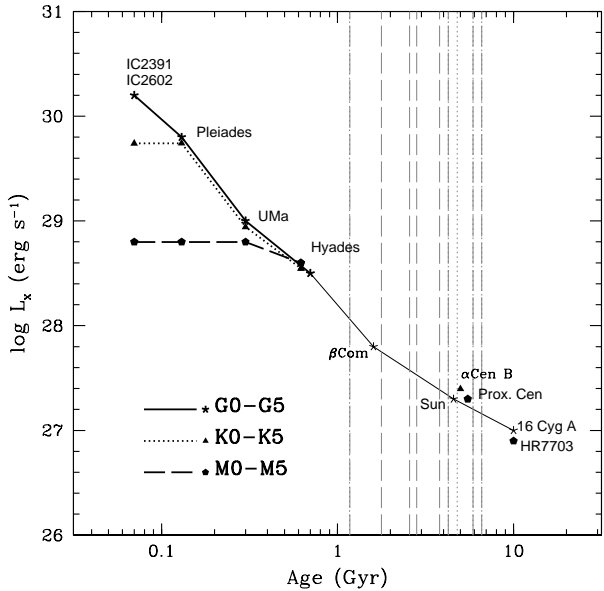}
\caption{$\log L_X$ vs. age diagram for G0-G5, K0-K5, and M0-M5 spectral type
bins. The vertical lines correspond to the ages of the WDs with reliable age
determinations from our study. The G0-G5 spectral type interval is already
fully covered from the \textit{The Sun in Time} project.}
\label{lx_age_new}
\end{figure}

Given the considerations above, we favour systems in which the contribution
of WD cooling time to the total age is at its maximum. In addition, we aim
at defining a sample of companion stars with a wide
range of spectral types that cover
as much of the important 1-6 Gyr age interval as possible.
From these two premises, we found that a constraint of $t_{\rm cool}/{\rm Age}$ $>$ 30$\%$
gives the best compromise between age robustness and sample size, so
therefore selected systems fulfilling this requirement. 
 These objects are 
highlighted in boldface in Table~\ref{results1}. Eighteen systems in our sample
fulfil this condition. Almost all of the cooler WDs analysed in
this work belong to this reliable sample, which is expected because their cooling
times are relatively long. The last column of Table~\ref{results1} includes, 
for reference, the $V-J$ index of the companion to each WD so that the proper
sample selection for the next step can be made. Following the same prescription
as was used in the {\em Sun in Time} project, one should define the spectral type (or
mass, or $V-J$) intervals and select stars that cover a wide age range. Given
the $V-J$ colours of our sample stars, we grouped them into five bins, namely
early K, late K, early M, early-mid M, and mid-late M. Table~\ref{spt_bins}
shows these bins with the $V-J$ and ages of the stars included in them. Each 
bin has at least two stars (to be added to the cluster averages
presented in $\S$2), covering roughly the interval between 1 and 5 Gyr. 
These are the stars for which we collect activity information so that 
they can be used to trace the time-evolution of their high-energy emissions
from the start of the zero-age main-sequence to at least the age of 
our Sun. This is illustrated in Fig.~\ref{lx_age_new}, which is the same as
Fig.~\ref{lx_age} but with vertical lines showing the position of the ages that
will be sampled with the stars we have analysed. The new ages cover
the relevant area that is currently unpopulated and beyond what can be studied
with open cluster data.  

\begin{table}[!t]
\caption{\textit{V-J} colour and age for stars with $t_{\rm cool}$ vs. total age
ratio $>$30\%, separated in spectral type bins and sorted by age within each
bin.}
\label{spt_bins}
\centering
\small
\begin{tabular}{lc}
\textit{V-J} & Age \\
\hline
\hline
\textit{Early K}&\\
1.46 & 1.17$_{-0.13}^{+0.14}$ \\
1.21 & 3.6$_{-0.9}^{+2.1}$ \\
1.90 & 5.0$_{-1.2}^{+2.5}$ \\
\hline
\textit{Late K}&\\
2.19 & 2.3$_{-0.7}^{+1.6}$ \\
2.47 & 5.4$_{-1.3}^{+2.6}$ \\
\hline
\textit{Early M}&\\
2.95 & 1.13$_{-0.07}^{+0.08}$ \\
2.97 & 1.17$_{-0.14}^{+0.15}$ \\
2.97 & 2.8$_{-0.8}^{+1.8}$ \\
3.13 & 4.3$_{-1.1}^{+2.3}$ \\
\hline
\end{tabular}
\hspace*{5mm}
\begin{tabular}{lc}
\textit{V-J} & Age \\
\hline
\hline
\textit{Early-mid M}&\\
3.85 & 1.58$_{-0.20}^{+0.18}$ \\
3.67 & 1.64$_{-0.20}^{+0.20}$ \\
3.21 & 3.9$_{-1.1}^{+2.6}$ \\
\hline
\textit{Mid-late M}&\\
4.18 & 1.58$_{-0.20}^{+0.27}$ \\
4.19 & 1.77$_{-0.35}^{+0.71}$ \\
4.08 & 2.6$_{-0.7}^{+1.7}$ \\
4.38 & 5.9$_{-1.4}^{+2.7}$ \\
4.31 & 6.7$_{-1.3}^{+2.8}$ \\
\hline
\\
\\
\end{tabular}
\end{table}

\section{Conclusions}

We studied a sample of 30 wide binary systems composed of a WD component
and a low-mass companion. We have obtained high-signal-to-noise low resolution
spectroscopy for 25 of the WDs, which allowed us to carry out a full analysis
of their spectra and a fit of WD models to derive atmospheric parameters,
$T_{\rm eff}$ and $\log g$. Total ages for the wide binary systems were
obtained from a procedure based on the cooling sequences of \cite{Sal00}, the
initial-final mass relationship of \cite{cat08b}, and the stellar tracks of
\cite{dom99}. We estimate that the total ages for 18 of the systems are 
especially reliable as the relative contribution from the cooling time of the
WD (which is supposed to be relatively unaffected by systematic errors) is greater
than 30\%. The resulting sample covers spectral types from early K to mid-late
M and, very importantly, it contains stars that populate the critical age
interval between 1 and 5 Gyr for which the determination of ages using other
methods is difficult. Most of our targets are within a few tens of
parsecs, so the study of their activity properties is much easier than
for stars belonging to intermediate- or old-age clusters. 

For the selected stars we are obtaining high-resolution spectroscopy to measure
chromospheric fluxes and X-ray data so that we can build reliable relationships
describing the time evolution of emissions related to stellar magnetic
activity. As our final goal,  our work should provide critical input
information to any effort of understanding the long-term evolution of the
atmospheres of planets around low-mass stars, including their potential
habitability.  It will also allow  better understanding of the evolution of
high-energy emissions from low-mass stars. The activity-age relationship will
allow us not only to address the time evolution of high energy radiation, but
also to use it as an age calibrator when the stellar high-energy emissions can
be measured. This age indicator is very interesting because it will be based on
the star's own properties and will be useful in a regime ($>1$ Gyr) where
estimating stellar ages is still very difficult.

\begin{acknowledgements}
We thank D. Koester for providing us with his white dwarf models. We
acknowledge support from the Spanish MICINN grant AYA2009-06934. S.C. is
supported by a Marie Curie Intra-European Fellowship within the 7th European
Community Framework Programme.
\end{acknowledgements}

\bibliographystyle{aa}
\bibliography{garces2}	
\clearpage

\end{document}